\documentclass{icrc29}
\usepackage{graphicx,amssymb,amsmath,times}
\begin{document}
\title[HESS\,J1825-137: A pulsar wind nebula ...]
{HESS\,J1825-137: A pulsar wind nebula associated with PSR B1823-13?}
\author[O.C. de Jager et al.] {O.C. de Jager$^a$, S. Funk$^b$, \& J. Hinton$^b$
        for the H.E.S.S. Collaboration  \\
        (a) Unit for Space Physics, North-West University, Potchefstroom, 
	    South Africa \\ 
        (b) Max-Planck-Institut f\"ur Kernphysik, Heidelberg, Germany \\
	}
\presenter{Presenter: O.C. de Jager (fskocdj@puk.ac.za), \  
saf-dejager-OC-abs1.og22-oral}

\maketitle

\begin{abstract}

HESS\,J1825--137 was detected with a significance of
8.1 $\sigma$ in the Galactic Plane survey conducted with the
H.E.S.S. instrument in 2004. Both HESS\,J1825--137 and the X-ray pulsar
wind nebula G18.0--0.7 (associated with the Vela-like pulsar PSR\,B1823--13)
are offset south of the pulsar, which may be the result of the SNR expanding
into an inhomogeneous medium. The TeV size ($\sim 35$ pc, for a distance of 4 kpc) 
is $\sim 6$ times larger than the X-ray size, which may be the result of 
propagation effects as a result of the longer lifetime of TeV emitting electrons,
compared to the relatively short lifetime of keV synchrotron emitting electrons. 
The TeV photon spectral index of $\sim 2.4$
can also be related to the extended PWN X-ray synchrotron photon index of $\sim 2.3$,
if this spectrum is dominated by synchrotron cooling.
The anomalously large size of the pulsar wind nebula can be explained
if the pulsar was born with a relatively large initial spindown power and
braking index $n\sim 2$, provided that the SNR expanded into the hot ISM
with relatively low density ($\sim 0.003$ cm$^{-3}$).

\end{abstract}

\section{Introduction}
PSR\,B1823--13 is a 101 ms evolved
pulsar with a spin-down age of $T=2.1 \times 10^{4}$ years
(for an assumed braking index of $n=3$)
\cite{Clifton} and in these properties very similar to the Vela
pulsar. It is located at a distance of $d=3.9\pm 0.4$
kpc \cite{Cordes_Lazio}. High resolution {\it XMM-Newton} observations
of the pulsar region confirmed a previous {\it ROSAT} detection of
an asymmetric diffuse nebula, which was hence given the name
G\,18.0--0.7 \cite{XMM}. For the compact core with extent
$R_{\mathrm{CN}}\sim 30''$ (CN: compact nebula) immediately
surrounding the pulsar, a photon index of
$\Gamma_{\mathrm{CN}}=1.6^{+0.1}_{-0.2}$ was measured with an
unabsorbed luminosity of $L_{\mathrm{CN}}\sim 9 d_4^2 \times 10^{32}$
erg\,s$^{-1}$ in the 0.5 to 10 keV range for a distance of $4d_4$
kpc. The corresponding pulsar wind shock radius is $R_s\leq 15''
= 0.3d_4$ pc. The compact core is embedded in a region of extended
diffuse emission which is clearly one-sided, revealing a structure
south of the pulsar, with an extension of $R_{\mathrm{EN}}\sim
5'$, (EN: extended nebula) whereas the $\sim 4'$ east-west
extension is symmetric around the north-south axis.  The spectrum of
this extended component is softer with a photon index of
$\Gamma_{\mathrm{EN}}\sim 2.3$, with an unabsorbed luminosity of
$L_{\mathrm{EN}}=3d_4^2\times 10^{33}$ erg\,s$^{-1}$ for the 0.5 to 10
keV interval. No associated supernova remnant (SNR) has been identified yet. 

\section{A possible association of HESS\,J1825--137, G18.0--0.7 and 3EG\,J1826--1302}
At $\gamma$-ray energies, PSR\,B1823--13 was proposed to power the
close-by unidentified EGRET source 3EG\,J1826--1302 \cite{EGRET}.
The region around PSR\,B1823--13 was observed as part of the survey of
the Galactic plane with the H.E.S.S. instrument \cite{HESSSCAN}. In
this survey, a source of very high-energy (VHE) $\gamma$-rays
(HESS\,J1825--137) 11$'$ south of the pulsar was discovered with a
significance of 8.1 $\sigma$. We note, that the new VHE $\gamma$-ray
source is located within the 95\% positional confidence level of the
EGRET source 3EG\,J1826--1302 and could therefore be related to this
as of yet unidentified object. More details of the H.E.S.S. observations, analyses
and merits of the possible association of HESS\,J1825-137 with G\,18.0--0.7 
was discussed by \cite{Aharonian_1825}.

Figure 1 shows a north-south declination slice of G18.0--0.7 in X-rays
and HESS\,J1825--137 in $\gamma$-rays. Both images show that the emission
peaks at the position of PSR\,B1823-13, but with extended emission shifted towards the south.
The prominent X-ray synchrotron peak at the pulsar position is as a result of the
compact X-ray nebula, where the magnetic field strength should be largest.
Whereas the extent of the X-ray emission is only $\sim 5$ arcmin, the
TeV emission is seen out to $\sim 0.5^\circ$.
Even though the position of the unpulsed EGRET source 3EG\,J1826--1302 is consistent
with that of HESS\,J1825--137 \cite{Aharonian_1825}, the poor angular resolution
of the EGRET instrument precludes a similar morphological study.

\begin{figure}[h]
\begin{center}
\includegraphics*[width=1.0\textwidth,angle=0,clip]{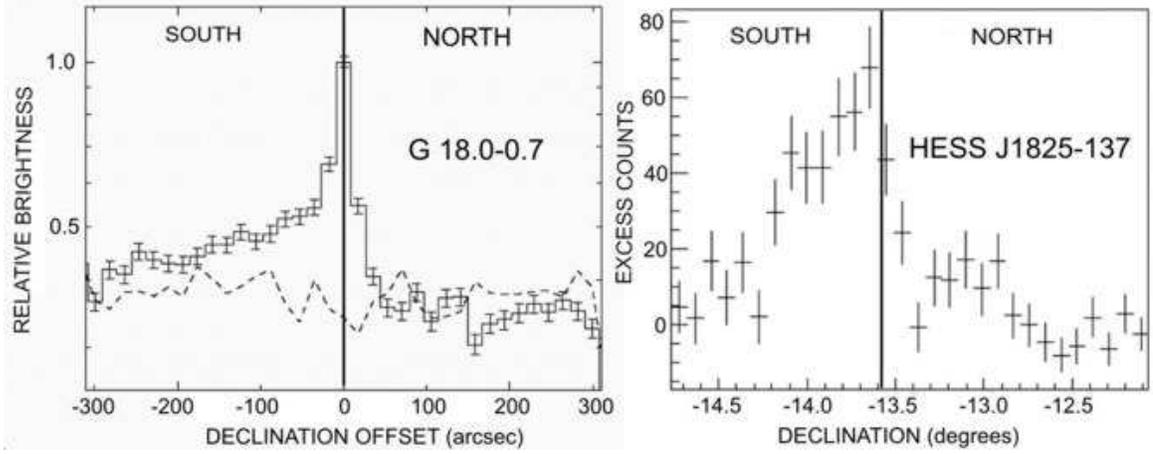}
\caption{\label {fig1} (a) left: north-south declination slice of
G\,18.0--0.7 in X-rays \cite{XMM} showing the bright compact nebular core
at the pulsar position and the $\sim 5'$ extended X-ray nebula
extending to the south of PSR\,B1823--13 (indicated by a vertical line). 
(b) right: similar slice for HESS\,J1825--137,
showing a similar extension south of the pulsar, but with a much larger size
of $\sim 0.5^\circ$.}
\end{center}
\end{figure}   

\section{The one sided nature of the PWN and the anomalous size of HESS\,J1825--137}
The one-sided nature of this PWN as seen in TeV was already suggested by \cite{XMM},
for the X-rays, based on the hydrodynamical simulations of \cite{Blondin} for Vela X
and earlier studies referenced by \cite{XMM}: It was assumed that
the density of the interstellar medium (ISM) around the progenitor
star was inhomogeneous along the north-south direction, with the
density/pressure towards the northern direction significantly larger relative
to the south. The reverse shock from the northern direction should then 
have crashed relatively early into the PWN, pushing the latter towards the south,
as observed.

The remaining problem with the association of HESS\,J1825--137 with G18.0--0.7/PSR\,B1823--13
is the anomalously large
size of the TeV source: the $0.5^\circ$ diameter of the source translates to a
radius of $R_{\mathrm PWN}\sim 17d_4$ pc, 
and in the absence of radiative cooling in a SNR shell, one would expect a SNR forward shock radius of 
$R_{\mathrm SNR}\sim 4R_{\mathrm PWN}=70d_4$ pc \cite{Swaluw}. 
One possible key towards solving this problem would be to 
consider a relatively large initial spindown power ($L_0$) and pulsar braking index less than
the canonical value of $n=3$, which would increase the age $T$ of the SNR.
Integrating the spindown power $E_{\mathrm PSR}=\int dt\, I\Omega\dot{\Omega}$ over the increased
age would also maximise the $PdV$ work done by the PWN against the SNR.
The Sedov-Taylor SNR size would also increase with increasing age (scaling as $T^{2/5}$), giving a radius of 
$$R_{\mathrm SNR}=78{\mathrm pc}~
\left(\frac{E_{\mathrm SN}}{10^{51}{\mathrm ergs}}\frac{0.003\,{\mathrm cm^{-3}}}{N}\right)^{0.2}
\left(\frac{1}{n-1}\right)^{0.4}.$$
Here $E_{\mathrm SN}\sim 10^{51}$ ergs is the SN explosion energy and $N\sim 0.003$ cm$^{-3}$
is the density of the hot phase of the ISM \cite{Ferriere}. A pulsar braking index
of $n=2$ (``LMC pulsar'' type) will increase the age of PSR\,B1823--13 to $\sim 42$ kyr
if we assume a relatively large $L_0$ (i.e. an initial spin period much less than 0.1 s).
This will imply a size consistent with the predicted size of $\sim 70$ pc.

Figure 2(a) shows the evolution of $E_{\mathrm PSR}$ with time for different values of
$n$ and $L_0$: it is also clear that the total PWN energy is maximised for minimum $n$ and
maximal value of $L_0$, but it is not yet clear how to relate this time dependent
spindown power to the evolution of the PWN in a quantitave way. 

\begin{figure}[h]
\begin{center}
\includegraphics*[width=1.0\textwidth,angle=0,clip]{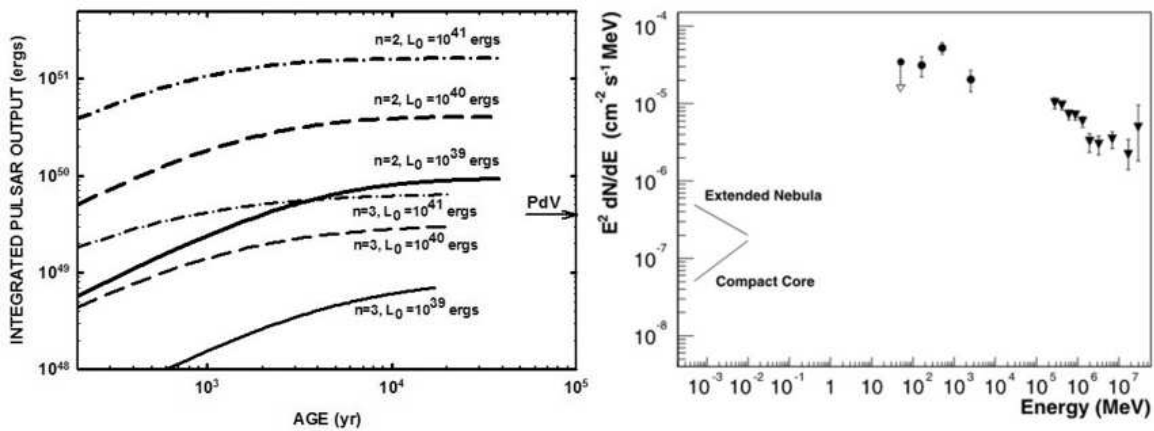}
\caption{\label {fig2} (a) left: integrated spindown power for PSR\,B1823--13
assuming different braking indices $n$ and initial spindown power $L_0$ values
as indicated. The arrow indicates the $PdV$ work required to displace
the observed PWN volume against a pressure of 5 eV/cm$^3$. 
The vertical position of the arrow would then scale linearly with pressure. 
(b) right: the spectral energy distribution
of the PWN of PSR\,B1823--13: HESS\,J1825--137 (triangles), G\,18.0--0.7 
(compact and extended nebula indicated by two lines)
and 3EG\,J1826--1302 (circles). For details see \cite{Aharonian_1825}.}
\end{center}
\end{figure}   

\section{The spectral energy distribution of the PWN of PSR\,B1823--13}
Aharonian et al.
\cite{Aharonian_1825} derived the spectrum of HESS\,J1825--137, which can
be represented by a power law in energy, with photon index 
$2.40\pm0.09_{stat}\pm0.2_{sys}$ and a flux above 230~GeV of
$(3.4\pm0.2_{stat}\pm0.8_{sys})\,\times\,10^{-11}$ cm$^{-2}$s$^{-1}$
(corresponding to 12\% of the Crab flux above that energy). This spectrum is shown
in Figure 2 (b), together with the X-ray spectra of G18.0--0.7 \cite{XMM} and 
3EG\,J1826--1302 \cite{Aharonian_1825}. It is clear that the TeV spectrum
of HESS\,J1825--137 does not extrapolate towards GeV energies, and a spectral flattening
below $\sim 10$ GeV is required. This spectral break may be the
result of a cooling effect, which is also seen in X-rays: 
The compact nebular X-ray photon spectral index of $\sim 1.6$ represents the uncooled
component, whereas the photon index of $\sim 2.3$ measured from the extended X-ray nebula,
represents the synchrotron cooled component (as a result of propagation effects). 
The TeV photon index of $\sim 2.4$ is then
consistent with inverse Compton scattering of these synchrotron cooled
electrons on the CMBR, provided that this cooled spectrum also
extends into the EUV/soft X-ray band, where electrons contribute
directly to the observed TeV emission for a field strength of $\sim 10\mu$G or less.

The expected intrinsic break in the injected spectrum, represented by the so-called
``$\Gamma$-factor'' of the wind at the pulsar wind shock of the compact nebula
may also contribute to the observed flattening: in this case the number spectrum of
electrons with energy $E_e>\Gamma m_ec^2$ is expected to follow an accelerated power law 
spectrum of the form $E_e^{-2.2}$, resulting in the observed (uncooled) compact nebular spectrum with a photon index of 1.6 (as discussed above).
Electrons with energy $E_e<\Gamma m_ec^2$ is then expected to have a much harder
spectrum. This component is however unconstrained by radio or optical observations. 

We also note that the energy flux in TeV is larger than the
synchrotron energy flux of the extended X-ray nebula. This dominance
of the TeV flux holds even if we extrapolate the spectrum of the extended
X-ray nebula to the EUV/soft X-ray band. The evaluation of this soft synchrotron
component is important, since electrons contributing to this soft unseen component, also
scatter CMBR photons into the H.E.S.S. band for field strengths of $\sim 10\mu$G or less:
Note that the {\it XMM} X-ray
spectrum is only measured within $5'$ from the pulsar, whereas electrons
contributing to this soft band should have a much longer lifetime. They can therefore
propagate to larger distances (as seen in TeV), resulting in a population of accumulated
electrons over the lifetime of the pulsar, whereas the X-rays seen in Figure 2 (b)
represent synchrotron emission from freshly injected electrons. This accumulation
effect should increase the intensity of the soft component accordingly. Thus, the
soft unseen synchrotron component should have a similar size as seen in TeV.
A test for this hypothesis would be to measure the size of G\,18.0--0.7
in the softest channels of XMM over an area similar to the TeV size. 
A more detailed study is however
required to model the intensity of this unseen soft synchrotron component. 

\section{Appendix: lifetimes for X-ray and TeV emitting electrons}
The respective synchrotron lifetimes of electrons in a transverse magnetic field of 
strength $B=10^{-5}B_{-5}$ G, radiating synchrotron photons with 
mean energy $E_{\mathrm keV}$ (in units of keV) and inverse Compton
scattered CMBR photons to a mean $\gamma$-ray energy of 
$E_{\mathrm TeV}$ (in units of TeV), are given by
$$\tau_{\mathrm X}=1.2\;{\mathrm kyr}B_{-5}^{-3/2}E_{\rm keV}^{-1/2}\;\;{\rm and}\;\;
\tau_\gamma=4.8\;{\mathrm kyr}B_{-5}^{-2}E_{\mathrm TeV}^{-1/2}.$$

\section{Acknowledgements}

We thank Felix Aharonian, Luke Drury \& Yves Gallant for useful discussions.


\begin{thebibliography}{99}
\bibitem{HESSSCAN}Aharonian, F.A. et al.,
({\it H.E.S.S. Collaboration}) 2005a, Science, 307, 1938
\bibitem{Aharonian_1825}Aharonian, F.A. et al.,
({\it H.E.S.S. Collaboration}) 2005b, in preparation
\bibitem{Blondin}Blondin, J.M., Chevalier, R.A.,
\& Frierson, D.M. 2001, ApJ, 563, 806
\bibitem{Clifton}Clifton, T.~R., Lyne, A.~G., Jones, A.~W., McKenna, J., \& 
Ashworth, M. 1992, MNRAS, 254, 177
\bibitem{Cordes_Lazio}Cordes, J.~M., \& Lazio,
T.~J.~W. 2002, preprint (Astro-ph/0207156)
\bibitem{EGRET}Nolan, P.~L, Tompkins, W.~F.,
Grenier, I.~A., \& Michelson, P.~F. 2003, ApJ., 597, 615
\bibitem{Ferriere}Ferriere, K. 1998, ApJ, 503, 700
\bibitem{XMM}Gaensler, B.~M., Schulz, N.~S.,
Kaspi, V.~M.  Pivovaroff, M.~J., \& Becker, W.~E. 2003, ApJ, 588, 441
\bibitem{Swaluw}van der Swaluw, E., Achterberg, A.,
Gallant, Y.~A., \& Toth, G. 2001, A\&A, 380, 309

\end{thebibliography}
\end{document}